# The AI Pentad, the CHARME$^2$D Model, and an Assessment of Current-State AI Regulation

Di Kevin Gao*†, Sudip Mittal†, Jiming Wu*, Hongwei Du*, Jingdao Chen†, Shahram Rahimi†
*California State University East Bay
†Mississippi State University
Email: kevin.gao, jiming.wu, hongwei.du@csueastbay.edu, sm3888, jdc1258, rahimi@msstate.edu

*Abstract*—Artificial Intelligence (AI) has made remarkable progress in the past few years with AI-enabled applications beginning to permeate every aspect of our society. Despite the widespread consensus on the need to regulate AI, there remains a lack of a unified approach to framing, developing, and assessing AI regulations. Many of the existing methods take a value-based approach, for example, accountability, fairness, free from bias, transparency, and trust. However, these methods often face challenges at the outset due to disagreements in academia over the subjective nature of these definitions. This paper aims to establish a unifying model for AI regulation from the perspective of core AI components. We first introduce the AI Pentad, which comprises the five essential components of AI: humans and organizations, algorithms, data, computing, and energy. We then review AI regulatory enablers, including AI registration and disclosure, AI monitoring, and AI enforcement mechanisms. Subsequently, we present the CHARME$^2$D Model to explore further the relationship between the AI Pentad and AI regulatory enablers. Finally, we apply the CHARME$^2$D model to assess AI regulatory efforts in the European Union (EU), China, the United Arab Emirates (UAE), the United Kingdom (UK), and the United States (US), highlighting their strengths, weaknesses, and gaps. This comparative evaluation offers insights for future legislative work in the AI domain.

## I. Introduction

AI has made remarkable progress in the past few years, with releases of DeepMind's AlphaGo, OpenAI's ChatGPT, Stability AI's Stable Diffusion occurring in accelerating successions. Rapid advancements in computing and improvements in algorithmic efficiency, coupled with data explosion from the digitized physical twins, are poised to remake human society [1].

Amid these significant technological breakthroughs and widespread excitement, there is also considerable societal uneasiness regarding AI's ethical and social challenges. These challenges include transparency and explainability, privacy protection, social justice and fairness, algocracy and human enfeeblement, and the potential for superintelligence [2].

As AI continues to evolve, a proactive and adaptive regulatory approach will be essential to harness its full potential while safeguarding the public interest. The European Union's Artificial Intelligence Act (AI Act) is a pioneering legislation aimed at regulating AI technologies within the EU. However, beyond the EU, considerable disagreement remains on how AI should be effectively regulated, with no universally accepted models to frame, construct, and evaluate AI regulations. Some existing methods adopt a value-based approach, emphasizing principles such as accountability, fairness, free from bias, transparency, and trust, as highlighted by Rêgo de Almeida et al. in their review of Artificial Intelligence regulation frameworks published between 2010 and 2020 [3]. However, these methods often face challenges at the outset due to disagreements in academia over the subjective nature of these definitions. This paper aims to address this gap by establishing a novel unifying model for AI regulation from the perspective of AI's most essential components, including human and organization, algorithm, data, compute, and energy.

We reviewed the literature on AI's essential components and found that Buchanan (2020) introduced the concept of the AI Triad, which comprises computing power, data, and algorithms [4]. Similarly, Mark Nitzberg and John Zysman proposed regulating AI based on the "fundamental drivers of the development and deployment of AI tools," which include algorithms, data, and the dominant Digital Platform Firms (DPFs) [5]. DPFs are defined as "entities that develop and manage the infrastructure and ecosystems where AI technologies, data, and algorithms are deployed, scaled, and utilized," such as Amazon, Google, Microsoft, and Meta.

While Buchanan's triad is insightful, it omits the critical human element. Humans have always been central to the AI revolution, shaping its development, deployment, and societal impact. Although AI can be considered a regulatory object [6], it is not a legal entity capable of assuming responsibility or liability. To ensure effective regulation, it is necessary to identify the ultimate benefactors and accountability holders within the AI ecosystem. For this reason, we propose incorporating human and organizational factors into the framework.

Nitzberg's concept of DPFs aligns with the human and organization component of our proposed AI Pentad, which expands upon the AI Triad by encompassing the social and organizational structures that influence AI's trajectory. From this perspective, the AI Pentad effectively integrates and extends Nitzberg's fundamental drivers.

The inclusion of energy as a component in the AI Pentad may face questions. However, we argue that energy is a crucial enabler of AI's scalability, sustainability, and measurability.

Further justifications for its inclusion will be provided in the following sections.

Additionally, we review AI regulatory enablers, including AI registration and disclosure, monitoring, and enforcement. We then introduce the CHARME²D Model, which examines the complex interplay between the AI Pentad and regulatory enablers. This model is subsequently applied to assess AI regulatory progress in selected countries and regions, including the EU, China, UAE, UK, and the US.

The contributions of this article are threefold: 1) the introduction of AI Pentad, an extension of the AI Triad, to better understand and identify regulatory intervention points within AI's core components; 2) the introduction of CHARME²D model, a universal framework that can help frame, construct, and evaluate legislative efforts; 3) a broad assessment of the AI regulatory progress of selected countries and regions against the CHARME²D model to highlight strengths, weaknesses, and gaps. This comparative evaluation offers insights for future legislative work in the AI domain.

## II. DEFINITIONS

*Artificial Intelligence* (AI) was coined in 1955 by John McCarthy, Marvin Minsky, Nathaniel Rochester, and Claude Shannon during the preparation of the Dartmouth Workshop . John McCarthy defined AI as "the science and engineering of making intelligent machines, especially intelligent computer programs. It is related to the similar task of using computers to understand human intelligence, but AI does not have to confine itself to methods that are biologically observable". [2]

For *regulation*, we take the Mitnick's definition: "the central element of the class of behaviors that might be termed 'regulation' is an interference of some sort in the activity subject to regulation – it is to be governed, altered, controlled, guided, regulated in some way". Mitnick further clarifies that this implies that regulated activities are not to be replaced or banned; they are only to be regulated. [7]

## III. AI PENTAD

In this section, we introduce the AI Pentad, a framework designed to provide a comprehensive understanding of AI and its ecosystem to effectively frame, structure, and evaluate AI regulations.

The AI Pentad consists of five elements: humans and organizations, algorithms, data, compute, and energy.

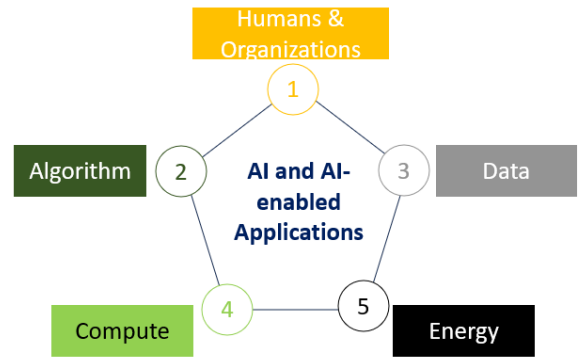

Fig. 1: The AI Pentad represents the five core elements of Artificial Intelligence: humans and organizations, algorithms, data, compute, and energy. It helps to methodically analyze elements in AI regulation.

### A. Humans and Organizations

Humans are AI's originators and ultimate benefactors, or sufferers if AI regulation fails. Organizations are the structures through which humans ethically and creatively govern and utilize AI. The human element is crucial across every aspect of AI including research, design, training, interpretation, monitoring, and supervision.

The AI Pentad starts with humans and organizations, acknowledging that human intentions, behaviors, and forced and unforced errors are atop all other factors in AI regulations.

### B. Algorithm

Algorithms define how data is processed and how decisions are made within AI applications. [8]. They are the engines that power AI, fundamental to its effectiveness and efficiency. They are critical because they provide the methods for solving problems, enabling machines to perform tasks that typically require human intelligence.

Advanced algorithms learn and adapt to new data and environments, continuously optimizing performance by adjusting underlying parameters to minimize errors and maximize accuracy.

### C. Data

Data serves as the foundation for designing, training, building, and refining AI models. During training, AI extracts relationships and recognizes patterns, while post-deployment, it makes inferences based on new data and learned models. Additionally, to provide personalized recommendations, AI relies on "data about user behavior and preferences to provide relevant suggestions". [8]

The success and reliability of AI systems are intrinsically linked to the quality and comprehensiveness of the data they utilize. Ensuring unbiased and comprehensive data is crucial to prevent skewed results and to uphold fairness.

### D. Compute

While data and algorithms have been integral to innovation for centuries, the transformative advancements in GPU

computing and the widespread availability of computational power through cloud providers and server farms have truly unlocked AI's potential. The substantial increase in computational resources has been pivotal for the development and success of modern AI systems. For example, training a model like LLaMA-65B requires an astounding 1,022,362 GPU hours using NVIDIA's A100-80GB system [9], a scale unimaginable just a few decades ago. At these billion-parameter scales, computational power has emerged as a scarce and valuable resource, highlighting its significance as a potential focal point for AI regulation.

In his 2020 paper, Buchanan cited OpenAI's study, which showed that compute drove AI progress more than data and algorithms from 2012 to 2018, challenging the traditional paradigm that innovation in AI is primarily driven by data and algorithmic improvements [4]. A recent technical report by Sastry et al., representing a collaborative effort between OpenAI, the Centre for the Governance of AI (GovAI), the Leverhulme Centre for the Future of Intelligence, and other prominent AI institutions, further underscores this point. The report identifies AI-relevant computational resources as an exceptionally effective point of intervention for AI regulation due to their detectability, exclusivity, quantifiability, and dependence on a highly concentrated supply chain [10].

*E. Energy*

Energy is an often-underappreciated yet increasingly critical component of AI development and operations. It plays a foundational role in powering the hardware and processes necessary for AI research. Systems that rely on deep learning and large-scale data processing are inherently energy-intensive. According to a report by Goldman Sachs, energy demand for data centers is projected to rise by 160% by 2030 [11], driven by a combination of slowing energy efficiency improvements and escalating AI energy requirements.

Leaders in the AI industry have underscored the vital role of energy in shaping the future of AI. At the Bosch Connected World conference, Elon Musk highlighted the escalating demand for energy infrastructure, stating, "A year ago, the shortage was chips; neural net chips. Then, it was very easy to predict that the next shortage will be voltage step-down transformers . . . and the next shortage will be electricity" [12]. Similarly, Mark Zuckerberg has publicly acknowledged that energy constraints have become the primary bottleneck in expanding AI data centers [13]. During the 2024 World Economic Forum in Davos, Sam Altman projected that breakthroughs in energy technology will be essential for the future of AI development [14]. Recognizing this, major technology companies, including Microsoft, Amazon, Google, and OpenAI, have made significant investments in energy infrastructure to secure reliable energy supplies [15] [16] [17] [18].

Energy plays a crucial role in AI, not only as a fundamental resource but also as a potential proxy for the computational complexity of AI services. In the context of large language models, for instance, different types of queries demand vastly different levels of computational effort and, consequently, energy consumption. A simple query like identifying the largest ocean in the world requires far fewer resources compared to generating a detailed two-page product description. Since AI services often scale with computational demands, energy consumption provides a direct, easy-to-understand, and measurable indicator of usage. This makes it a viable basis for pricing AI services, offering a clear, consumption-driven model.

Energy supply, which falls under the purview of government regulation, represents a key lever for policymakers to influence AI development. Governments have the authority to approve, restrict, or modify energy infrastructure projects, and this regulatory control could be pivotal as energy emerges as a critical constraint, akin to the scarcity of AI talent, data availability, and the shortage of Nvidia GPUs. We posit that energy — the fundamental resource powering AI systems — will become a decisive factor in determining the trajectory of AI advancement. As AI systems grow increasingly energy-intensive, addressing energy availability and efficiency will not only mitigate constraints but also serve as a strategic focal point for regulatory intervention.

## IV. AI REGULATORY ENABLERS

AI regulatory enablers refer to the mechanisms that regulatory bodies utilize to implement and enforce regulations effectively. These enablers typically encompass three primary forms: AI registration and disclosure, AI monitoring, and AI enforcement.

This framework has been widely applied across various regulatory domains. For example, the U.S. Food and Drug Administration (FDA) employs similar mechanisms, including licensing (registration), inspections (monitoring), and enforcement actions to ensure compliance with health-related product standards. Similarly, the U.S. Environmental Protection Agency (EPA) highlights the importance of regulation, monitoring, and enforcement in shaping environmental behavior. The International Energy Agency (IEA) further emphasizes the significance of monitoring, verification, and enforcement in the implementation of energy efficiency standards and labeling programs.

In the context of AI regulation, Anderljung et al., in their article "Frontier AI Regulation: Managing Emerging Risks to Public Safety," proposed a framework that includes encouraging voluntary self-regulation and certification, granting regulators powers to detect violations and issue penalties for non-compliance, and requiring licenses for the development and/or deployment of frontier AI systems [19]. This framework aligns closely with the principles of registration, monitoring, and enforcement. Similarly, Ferrari et al., in their article "Observe, Inspect, Modify: Three Conditions for Generative AI Governance," discuss a governance model comprising observation, inspection, and modification, which mirrors the foundational framework of registration, monitoring, and enforcement [20]. These examples demonstrate the broad applicability of this tripartite regulatory approach in managing the risks and complexities associated with AI technologies.

Here are more detailed descriptions:

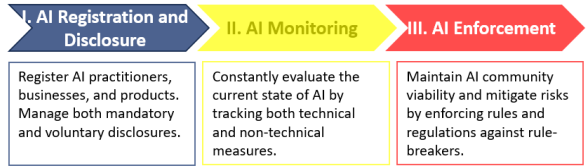

Fig. 2: AI Regulatory enablers come in three main forms: registration and disclosure, monitoring, and enforcement.

- **AI Registration and Disclosure:** This encompasses the registration of AI practitioners, businesses, and products. Practitioner registration includes individuals involved in AI ideation, research, design, development, training, testing, and deployment, as well as key managerial personnel. Business registration applies to both for-profit and non-profit organizations.
- **AI Monitoring:** Continuous evaluation and oversight of AI systems to ensure compliance with regulatory standards. This includes tracking technical and non-technical measures to identify potential risks, biases, and errors.
- **AI Enforcement:** Mechanisms to ensure adherence to established laws, including compliance checks, audits, and inspections to verify that AI systems operate within legal and ethical boundaries.

## V. THE CHARME$^2$D MODEL

The CHARME$^2$D Model combines AI regulatory enablers (left) and the AI Pentad (right). The model derives its name from the initial letters of the components of the AI Pentad and the regulatory enablers.

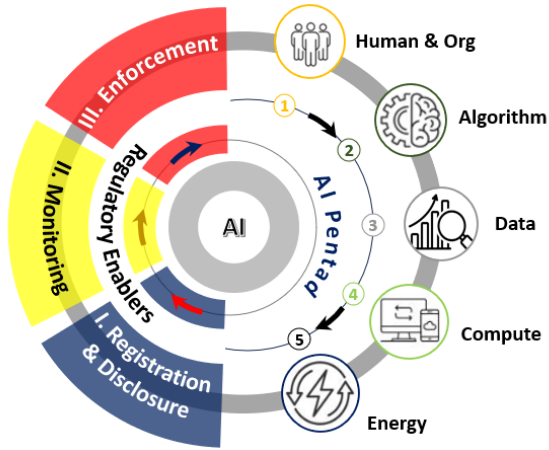

Fig. 3: The CHARME$^2$D Model

The concept behind the CHARME$^2$D model is straightforward: for each component of the AI Pentad, regulations can be "enabled" by leveraging the forces of each enabler. For instance, the algorithm component includes aspects of algorithm registration and disclosure, algorithm monitoring, and enforcement actions related to algorithms. This approach applies similarly to the other components of the AI Pentad.

When framing, constructing, or assessing AI regulations for a state, country, or other political entity, the CHARME$^2$D model can be utilized to evaluate their completeness, strengths, and weaknesses.

A more complex form of the CHARME$^2$D model is presented in a two-dimensional format, as illustrated in Fig. 4, where the combinations of the AI regulatory enablers and the AI Pentad are fully expanded.

## VI. EVALUATING AI REGULATORY EFFORTS OF SELECTED COUNTRIES AGAINST THE CHARME$^2$D MODEL

In this section, we will assess AI regulatory approaches and progress in the EU, China, UAE, UK, and the US to identify strengths, weaknesses, and key gaps. This comparative evaluation is valuable in offering key insights for future legislative efforts in the AI domain. This assessment is up-to-date as of June 30, 2024, and only federal laws, national laws, and supranational laws are considered.

### A. EU: Horizontal AI law anchored on risks

The EU's AI Act (AIA) is the world's first comprehensive AI regulation. It entered into force on August 1, 2024. EU AIA is a risk-based, sector-agnostic, and horizontal law designed to cover all AI-enabled applications. EU's focus has been risk containment and consumer protection.

- **Human and Org** The implementation of the AI Act is delegated to the EU AI Office. EU Member States will establish or designate at least one notifying authority and at least one market surveillance authority [21]. There are currently no registration or disclosure requirements for AI practitioners or AI businesses.
- **Algorithm** Harmful, abusive, or dangerous AI systems are prohibited. High-risk AI systems must be registered in the EU database, comply with specific requirements, and demonstrate compliance upon request. Limited-risk AI systems are subject to transparency obligations and disclosure requirements. Minimal-risk AI systems have no mandatory requirements but may adhere to a voluntary code of conduct. The AI Act provides exemptions for AI systems used exclusively for military, defense, national security, or solely for scientific research and development.[21]
- **Data** in the EU is governed by the 2018 General Data Protection Regulation (GDPR).
- **Compute**: General Purpose AI models, whose cumulative amount of computation used for training exceeds $10^{25}$ floating point operations, are presumed to be systems with systemic risk. These systems are subject to additional model evaluation, risk assessment, incident reporting, and cybersecurity requirements.
- **Energy** The EU does not regulate AI-related energy use.

The following diagram summarizes EU's current regulatory efforts against the CHARME$^2$D Model.

## Fig. 4: CHARME²D Model in Table Format

| AI Pentad | AI Regulatory Enablers | | |
|---|---|---|---|
| | **I. Registration & Disclosure** | **II. Monitoring** | **III. Enforcement** |
| ① Humans & Orgs | Register AI practitioners, AI businesses (for-profit and nonprofit), as well as service providers with proper disclosure. | Monitor the certification and annual affirmation of ethical standards. Monitor mergers and acquisitions. | Enforce member disciplinary actions, suspend and ban membership, and prosecute criminal actions in courts. |
| ② Algorithm | Register instances of AI algorithms with mandatory and optional disclosures of new capabilities, limitations, and risks. | Monitor algorithm usage, incidents, complaints, and other abnormal activities and investigate. | Ban certain harmful algorithms. Limit usage of some frontier models. Continuously enforce algorithmic safety rules. |
| ③ Data | Register data usage by AI product companies including data sourcing, data storage, and data divestiture. | Monitor data usage exceptions and investigate. Observe the rights of the data owners. | Settle data usage disputes; ban unauthorized data usage; penalize violators. |
| ④ Compute | Register on-prem and cloud computing by providers and users. Manage chip registry. | Monitor abnormal consumption of computing power. | Bar or limit violators or unqualified users from acquiring certain grades of chips or certain levels of computing power. |
| ⑤ Energy | Register planned energy usage for both on-grid and off-grid for training and inferencing. | Monitor anomalies in energy usage and investigate. | Bar or limit violators/unqualified players from certain energy consumption levels/intensities. |

## Fig. 5: EU AI regulation against the CHARME²D Model.

| AI Pentad | I. Registration and Disclosure | II. Monitoring | III. Enforcement |
|---|---|---|---|
| ① Human & Org | Basic | Unaddressed | Unaddressed |
| ② Algorithm | Adequate | Basic | Basic |
| ③ Data | Adequate | Adequate | Adequate |
| ④ Compute | Very Basic | Unaddressed | Unaddressed |
| ⑤ Energy | Unaddressed | Unaddressed | Unaddressed |

*B. China: Vertical and algorithm-centric AI regulation, but going horizontal*

China's AI regulatory approach is primarily vertical, focusing on algorithms, but it is becoming more comprehensive and horizontal. China's priority has been technology supremacy and economic benefits with some balance in consumer protection.

China has changed regulatory directions a few times in the last few years, including initial industry self-regulation (2017-2020) with AI Industry Alliance issued the AI Industry Self-Regulation Convention [22], and following light regulatory oversight (2020-2022) with the issuance of Guidelines to the Construction of the National New-Generation AI Standard System [23], to an increasingly techno-specific and mandatory regulation after 2022. Since 2022, China has enacted several national laws, including Internet Information Service Algorithmic Recommendation Management Provisions that took effect on March 1, 2022 [24], the Provisions on the Administration of Deep Synthesis Internet Information Services that took effect Jan 1, 2023 [25], and the Interim Measures for the Management of Generative Artificial Intelligence Services that took effect Aug. 15, 2023 which "encourages the innovative use of generative AI in all industries and fields". [26]

- **Human and Org** The Cyberspace Administration of China (CAC) is currently leading AI governance, though AI regulation may extend beyond CAC's core competency of online content controls. The Ministry of Science and Technology is another key player. There are currently no registration and disclosure requirements for AI practitioners and AI businesses.
- **Algorithm** Chinese laws have specific and sectorial requirements for algorithms:
  - The Internet Information Service Algorithmic Recommendation Management Provisions prohibits excessive price discrimination and protects the rights of workers subjected to algorithmic scheduling.
  - Deep Synthesis Regulation mandates conspicuous labels on synthetically generated content.
  - Interim Generative AI Regulation requires both the training data and model outputs to enhance "factualness and accuracy". [26]

  Based on the new legislation, developers are required to file information with China's algorithm registry on how algorithms are trained and pass a security self-assessment.
- **Data** Governed by the Data Security Law and the Personal Information Protection Law.
- **Compute** Chinese AI laws currently do not regulate based on compute capacity.
- **Energy** Chinese AI laws currently do not regulate AI-

related energy usage.

The following diagram summarizes China's current regulatory efforts against the CHARME$^2$D Model.

Fig. 6: China AI regulation against the CHARME$^2$D Model.

### C. United Arab Emirates (UAE): Active regulators but fragmented efforts

Mainland UAE (excluding the Financial Free Zones) currently has no comprehensive national laws for AI. Instead, it has been relying on guidelines and decrees. UAE's focus has been on technological advancement and economic benefits. Here are a few major legislative works in AI regulation:

- In 2017, UAE has a dedicated "AI Office" and appointed a Minister of State for AI, Digital Economy, and Remote Work Applications to set strategic directions and exercise regulatory oversight. [27]
- The AI authorities have published a series of non-binding national guidelines, including the 2021 DeepFake Guide [28], the 2022 AI Ethics Guide [29], and the 2023 AI Adoption Guideline in Government Services [30].

The evaluation of AI regulation in the UAE against the CHARME$^2$D model is as follows:

- **Human and Org** Currently, the AI Office and Minister of AI exercise regulatory oversight on AI. There are no requirements for practitioner registration or AI business registration.
- **Algorithm** No laws regulating algorithm.
- **Data** Governed by the Data Protection Law of 2021.
- **Compute** No laws regulating compute.
- **Energy** No laws regulating energy.

Measuring against the CHARME$^2$D model, UAE's AI regulation is still primitive.

### D. UK: Non-statutory approach to instigate innovations

The United Kingdom's approach to AI regulation is pro-innovation, non-statutory, and favors the application of existing laws over introducing new comprehensive regulations. UK's focus has been innovation and economic development while balancing risk containment and consumer protection.

On August 3, 2023, the UK government issued the "AI Regulation White Paper", outlining cross-sectoral principles for existing regulators to interpret and apply within their remits to drive safe, responsible AI innovation.

On February 6, 2024, the UK Government provided a written response to the feedback on the White Paper. It clarified that the UK "will not put these principles on a statutory footing initially. New rigid and onerous legislative requirements on businesses could hold back AI innovation and reduce our ability to respond quickly and in a proportionate way to future technological advances. Instead, the principles will be issued on a non-statutory basis and implemented by existing regulators." [31]

The evaluation of AI regulation in the UK against the CHARME$^2$D model is as follows:

- **Human and Org** UK currently has no central AI agencies coordinating AI regulatory responses. There are currently no AI practitioner or business registration requirements.
- **Algorithm** The UK government differentiates three types of the most powerful AI systems: highly capable General Purpose AI (GPAI), highly capable narrow AI, and agentic AI. Rules for algorithms are being developed.
- **Data** Governed by the Data Protection Act of 2018.
- **Compute** The UK does not regulate AI using compute resources.
- **Energy** The UK does not regulate AI-related energy usage.

Overall, the UK AI regulation framework is still in its early stages when assessed against the CHARME$^2$D model.

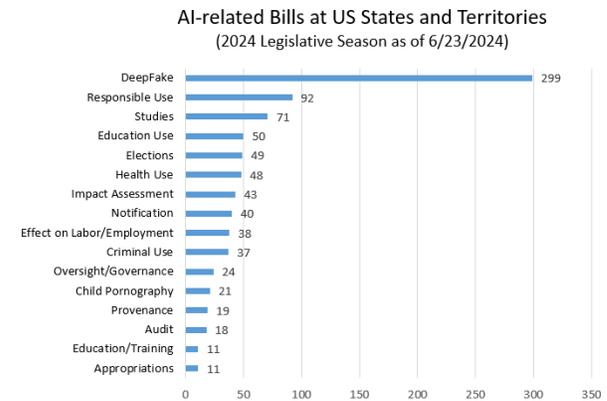

Fig. 7: The US states took an active role in regulating AI. In the 2024 legislative year till June 23, 2024, US states have 1,178 AI bills, excluding autonomous vehicle-related bills.

### E. United States: Congress takes a backseat as the administration and the states take center stage.

Currently, the US has no nation-wide laws on AI regulation. It relies on expanding different departments' existing mandate to cover AI. From that perspective, the US's AI regulatory approach is sectorial. For example, the Security and Exchange Commission (SEC) would expand oversight of anything AI related to securities and the Energy Department would expand its oversight of anything AI related to energy.

US's focus has been technology supremacy and economic opportunities while leveraging current laws for risk containment and consumer protection.

The US Congress has shown strong interest in AI regulation. However, progress was slow. In the 2024 legislative season, Senators Romney, Reed, Moran, and King unveiled a framework to mitigate extreme AI risks, while Senator Cantwell introduced bill S.4178, the Future of Artificial Intelligence Innovation Act of 2024. Additionally, Senate Majority Leader Chuck Schumer revealed a roadmap for Artificial Intelligence policy alongside a group of bipartisan senators. Despite this strong bipartisan interest in AI regulation, the prospects of passing comprehensive AI legislation remain uncertain.

In the absence of progress in Congress, the Biden Administration has utilized Executive Orders to address the regulatory gap. On October 30, 2023, President Biden issued Executive Order (EO) 14110, titled "Safe, Secure, and Trustworthy Development and Use of Artificial Intelligence." Previously, on December 8, 2020, President Biden issued Executive Order (EO) 13960, "Promoting the Use of Trustworthy Artificial Intelligence in the Federal Government." While EO 13960 focuses on AI applications within the Federal Government, EO 14110 applies to all industries. However, Executive Orders can be invalidated by future presidents and are not a replacement for laws passed by Congress.

States have also increased their efforts to regulate AI. According to the National Conference of State Legislatures, as of June 23, 2024, US states and territories have introduced 1,178 AI bills in the 2024 legislative season, not including bills related to autonomous vehicles [32].

- **Human and Org** There is no central agency coordinating AI regulation. Currently, there are no practitioner registration or business registration requirements for AI.
- **Algorithm** There are no current regulatory requirements for algorithms.
- **Data** The US does not have an updated federal data law. It relies on the 1974 US Privacy Act, HIPAA, COPPA, and state laws such as the California Privacy Rights Act (CPRA) of 2020.
- **Compute** According to Biden's Executive Order (EO) 14110, reporting is required for:
  - any model that was trained using a quantity of computing power greater than $10^{26}$ floating-point operations (FLOP), or using primarily biological sequence data and using a quantity of computing power greater than $10^{23}$ FLOP.
  - any computing cluster that has a set of machines physically co-located in a single datacenter, transitively connected by data center networking of over 100 Gbit/s, and having a theoretical maximum computing capacity of $10^{20}$ FLOP per second for training AI.
- **Energy** The US currently does not regulate AI-related energy usage.

The following diagram summarizes US's current regulatory efforts against the CHARME$^2$D Model.

Fig. 8: US AI regulation against the CHARME$^2$D Model.

Overall, all jurisdictions exhibit weaknesses in the Human and Organization category. Data regulation appears to be the most mature area. Algorithm regulation is emerging as a key focus for regulators, although categorizations, risk assessments, and sector-specific applications are still being refined. Compute regulation is also showing signs of becoming a significant area for regulatory intervention. The regulation of energy usage in AI remains largely undeveloped.

## VII. CONCLUSION

In this article, we introduced the AI Pentad, the AI regulatory enablers, and the CHARME$^2$D framework as a model to assess progress in AI regulation. We then evaluated the AI regulatory efforts of selected countries and regions against the CHARME$^2$D model. This comparative evaluation is valuable in offering key insights for future legislative efforts in the AI domain. Our analysis concluded that current AI regulation is still in its nascent stage. While data regulation is relatively mature, algorithm regulation has just commenced, compute regulation is still scant, regulation of humans and organizations is minimal, and energy regulation remains largely unaddressed. We recommend that regulators deploy the novel CHARME$^2$D model to frame, construct, and evaluate future AI regulations. AI regulation is a rapidly evolving field. The data utilized in this study is current as of June 30, 2024; however, more recent developments may not be captured.

## VIII. ACKNOWLEDGEMENTS

The work reported herein was supported by the National Science Foundation (NSF) (Award #DGE-2246920). Any opinions, findings, conclusions or recommendations expressed in this material are those of the authors and do not necessarily reflect the views of the NSF.